# Effective flows across diffusio-phoretic membranes

K. Wittkowski[1], P.G. Ledda[3], E.C. Giordano[2], F. Gallaire[1], and G.A. Zampogna*[2]

[1]LFMI, École Polytechnique Fédérale de Lausanne, CH-1015 Lausanne, Switzerland
[2]DICCA, Università degli Studi di Genova, Via Montallegro 1, 16145 Genova, Italy
[3]DICAAR, Università degli Studi di Cagliari, Via Marengo 2, 09123 Cagliari, Italy

## Abstract

Flows enabled by phoretic mechanisms are of significant interest in several biological and biomedical processes, such as bacterial motion and targeted drug delivery. Here, we develop a homogenization-based macroscopic boundary condition which describes the effective flow across a diffusiophoretic microstructured membrane, where the interaction between the membrane walls and the solute particles is modeled via a potential-approach. We consider two cases where potential variations occur (i) at the pore scale and (ii) only in the close vicinity of the boundary, enabling a simplified version of the macroscopic flow description, in the latter case. Chemical interactions at the microscale are rigorously upscaled to macroscopic phoretic solvent velocity and solute flux contributions, and added to the classical permeability and diffusivity properties of the membrane. These properties stem from the solution of Stokes-advection-diffusion problems at the microscale, some of them forced by an interaction potential term. Eventually, we show an application of the macroscopic model to develop minimal phoretic pumps, showcasing its suitability for efficient design and optimization procedures.

## 1 Introduction

Flows induced by chemical and electrical interactions between solid boundaries and a surrounding fluid are frequently encountered in nature as well as in engineering applications. Concentration gradients of different species can act as the driving force for chemical and biological transport: bacteria motion is often induced by concentration differences [Adler, 1975]. Plants heavily rely on chemical gradients to ensure water and sugar transport [Bohr et al., 2018]. Inspired by nature, artificial swimmers generate the concentration gradients required to propel themselves, e.g., through chemical reactions catalyzed at their surface [Anderson, 1989, Michelin et al., 2015, Paxton et al., 2004, Palacci et al., 2013, Brady, 2021, Peng et al., 2022], so-called *phoresis*. Particles may self-propel thanks to asymmetric coatings on their surface. These exhibit initial directional self-motion while, at large times, random walk paths emerge [Howse et al., 2007]. However, particles dispersed in colloidal dense suspensions can be effectively driven thanks to concentration gradients, generating steady clusters [Theurkauff et al., 2012]. In addition, temporal variations of the solute concentration can be exploited to drive particles' segregation and spatial patterns [Palacci et al., 2010]. Conversely, concentration gradients at fixed boundaries can be employed to induce so-called *phoretic* flows. As an example, the retinal pigmental endothelium (RPE) pump exploits concentration gradients to transport ions and molecules from the retina to the choroid and *vice versa* through the external layer of the vitreous, with the final result of draining excess fluid from the subretinal space [Sharma and Ehinger, 2003, la Cour and Tezel, 2005, Dvoriashyna et al., 2018, 2020].

Controlled flow manipulation at the micro- or nanoscale through minimal pumps recently found application also in microfluidics, including many applications in the field of biological analysis and screening [Whitesides, 2006]. Promising applications also include energy harvesting through the synergy of evaporation mechanisms with transport induced by electro-chemical interactions with nanopillar arrays, so-called *hydrovoltaic* effect [Anwar and Tagliabue, 2024], enabled by a concentration gradient of ions. Another example is the development of minimal phoretic pumps through the coating (so-called *phoretic* layer to recall the interaction with the concentration gradient) of the inner walls of a microfluidic channel [Michelin and Lauga, 2019]. The authors modeled the interaction with the phoretic layer as a non-zero surface tangential velocity, which is proportional to the concentration gradient along the normal direction through the so-called *local phoretic mobility*. Yu et al. [2020] employed similar phoretic-slip structures for the same purpose. As a matter of fact, it has been shown that tailored phoretic membranes, fully permeable to the solute and composed of an array of micro or nanochannels, can effectively drive a flow thanks to the presence of a concentration gradient across the structure [Lee et al., 2014]: this approach would potentially circumvent technical challenges related to the fabrication of selective membranes based on size exclusion [Werber et al., 2016]. Within this perspective, a rigorous derivation of the flow across fully permeable phoretic membranes may be useful to enhance predictability as well as promote efficient procedures for rational design and, eventually, facilitate their deployment within different contexts.

---

*Corresponding author e-mail: giuseppe.zampogna@unige.it

A first theoretical framework for the description of diffusio-phoretic flows can be identified in Derjaguin et al. [1947]. By leveraging hydrodynamic and thermodynamic considerations, the flux of a given chemical species in the vicinity of a phoretic surface is related to pressure and chemical potential differences. This result stems from a set of Stokes-Smoluchowski equations with a potential source term, which are nowadays the state of the art in the continuum modelling of interfacial phenomena [Kirby, 2010]. The reactive materials employed in these phenomena often present intricate multi-scale structures involving pores [Yoshida et al., 2017, Herman and Segev, 2024], pillars [Yu et al., 2020], and Janus coatings [Yang et al., 2016]. The numerical description of these structures, likewise porous membranes [Chen, 2021] or porous nano-carriers [Li et al., 2022], may be prohibitive from a computational point of view, because of the large dispersion of length and time scales within the same problem. In addition, results can be hardly generalized beyond the specific structure under consideration, and thus of little use in industrial applications.

Several works develop simplified models valid at the scale of the whole membrane.

The description of biological and engineered phoretic and osmotic flows typically employs ad-hoc empirical laws whose foundations lie in the Spiegler-Kedem-Katchalsky equations [Kedem and Katchalsky, 1958, Spiegler and Kedem, 1966]. These equations quantify the flow rates across the membrane via a linear combination of the contributions related to solvent pressure and solute concentration differences. These models are often developed for specific membrane geometries [Saffman, 1960, Malone et al., 1974], or depend on empirical parameters [Koter, 2006], with subsequent difficulties in the generalization of these results and the necessity of finely tuned experimental measurements to estimate permeability, slip and equivalent diffusivities [Peeters et al., 1998].

More sophisticated attempts tried to upscale a molecular description to the whole structure scale [Cardoso and Cartwright, 2014]. Averaging techniques [Whitaker, 1998] describe flows through reactive bulk porous media, based on Stokes-Smoluchowski equations containing a potential source term, valid within each pore [Wood et al., 2004, Veran et al., 2009]. Within this context, homogenization theory is a suitable tool to describe flows through permeable structures such as porous media [Mei and Vernescu, 2010] and microstructured boundaries [Jiménez Bolaños and Vernescu, 2017]. Recently, Zampogna and Gallaire [2020] introduced a homogenization-based approach to quantify the flow across thin microstructured permeable membranes. The distinct separation between the macroscopic size of the membrane and the microscopic size of its microstructure enables a mathematical description that scales up the microscopic flow characteristics to the observable macroscopic level. Homogenization theory establishes a rigorous connection between macroscale and microscale flows through characteristic problems solved at the microscale, to determine permeability and filtration properties. This methodology has been used to describe hydrodynamic flows coupled with passive chemical transport in microstructured membranes, assuming negligible inertia within pores [Zampogna et al., 2022, 2023]. The predictive accuracy of homogenization theory for thin membranes was also tested in Ledda et al. [2021], who presented an approach to optimize flows through microstructured membranes. In these works, the hydrodynamic flow for the solvent equation obeys a stress-jump condition where the velocities at each side of the membrane are proportional to the stresses acting on the membrane, at the macroscopic scale. Similarly, the concentration field presents a flux-jump due to the chemical interactions with the microstructure. However, these models still lack the coupling between the hydrodynamic flow and active chemical transport at the solute-solvent-solid wall boundaries.

Here, we propose a homogenization-based boundary condition to describe flows through microstructured phoretic membranes, as in figure 1a. The outcome of the procedure, whose workflow is sketched in figure 1b, is a macroscopic condition that must be imposed over a thin homogeneous interface between two fluid regions (red smooth surface in figure 1c). This model also accounts for interactions with the solid boundaries that compose the microstructure, via an interaction potential that models the near-wall interactions driving the fluid motion through the pores of the membrane. To give a comprehensive view of the problem, we distinguish two cases: case (i) in which the interaction potential shows appreciable variations within the microscopic cell (so-called *long-range* potential) and case (ii) where the characteristic variation of the interaction potential is much smaller than the size of pores so that chemical interactions can be described as a boundary condition on the solid walls (so-called *short-range* potential). The paper is organized as follows: Section 2 presents the governing equations at the pore scale. Section 3.1 is devoted to the derivation of the homogenization-based boundary condition, for the long-range potential case, whose microscale results are presented and validated within a macroscopic context in Section 3.2. Section 4 is analogous to Section 3, but focused on the short-range potential case. Section 4 shows how phoretic properties of membranes can tune the flow features surrounding the membrane itself. The work is concluded with perspectives, limitations and possible next developments to overcome these limitations.

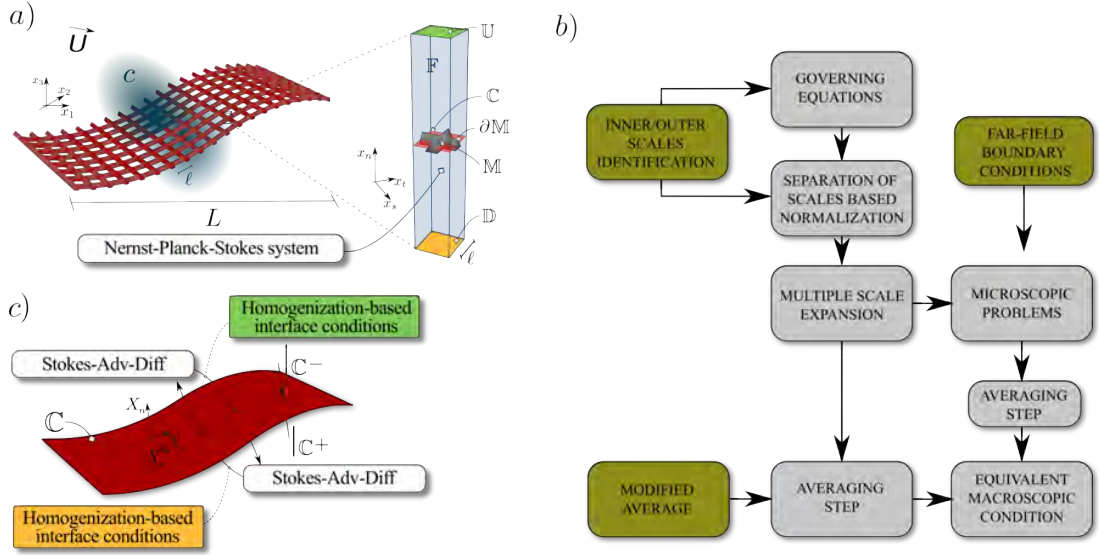

Figure 1: Fluid flow encountering a thin permeable membrane. The fluid is a two-component mixture of a solvent and a solute. Frame $a$): Real full-scale membrane realized by the periodic repetition of the microscopic cell represented on the right. A geometrical analysis of the membrane identifies two antipodal length scales; $L$ represents the size of the whole membrane, while $\ell$ is the typical size of the microscopic periodic structure. Frame $b$): diagram of the procedure used to deduce the macroscopic interface condition to describe interfacial flows through microstructured permeable walls, initially developed in Zampogna and Gallaire [2020]. Frame $c$): From a macroscopic point of view the membrane, denoted with $\mathbb{C}$, corresponds to a fictitious interface between two bulk fluid regions, provided with a local frame of reference $(X_s, X_t, X_n)$. We define the upward membrane side ($\mathbb{C}^-$) as the side of $\mathbb{C}$ whose outer normal coincides with $X_n$ in the sketch, counter imposed on the downward opposite side ($\mathbb{C}^+$).

## 2 Phoretic flows near solid boundaries

We consider the viscous, incompressible flow through a microstructured membrane of a diluted solute of diffusivity $D$ transported by a Newtonian fluid of density $\rho$ and dynamic viscosity $\mu$ (the *solvent*). The dimensional solute concentration and solvent flow fields (velocity and pressure) are denoted as $\hat{c}$ and $(\hat{u}_i, \hat{p})$, respectively. These quantities are defined within the fluid domain $\mathbb{F}$ (see figure 1), of characteristic length $\ell$, while the size of the membrane is denoted as $L$. The ratio between these two quantities defines the separation of scales parameter $\epsilon = \ell/L \ll 1$. We introduce the non-dimensional interaction potential $\phi$, modelling the chemical interactions between the solid boundaries and the solute particles, the Boltzmann constant $k_b$ and the absolute temperature of the system T. The Navier-Stokes equations are coupled with Fick's law for the conservation of the solute flux and both equations contain a potential source term. The set of governing equations read

$$\rho\left(\hat{\partial}_\tau \hat{u}_i + \hat{u}_j \hat{\partial}_j \hat{u}_i\right) = -\hat{\partial}_i \hat{p} + \mu \hat{\partial}^2_{ll} \hat{u}_i - \hat{c}\hat{\partial}_i \hat{\phi}, \quad \hat{\partial}_i \hat{u}_i = 0, \quad (1)$$

$$\hat{\partial}_\tau \hat{c} + \hat{\partial}_i \hat{F}_i = 0, \quad (2)$$

where $\hat{F}_i$ is the concentration flux

$$\hat{F}_i = \left(\hat{u}_i + \hat{u}_i^P\right)\hat{c} - D\hat{\partial}_i \hat{c} = \hat{u}_i \hat{c} + D\left(\frac{\hat{c}}{k_b \text{T}}\hat{\partial}_i \hat{\phi} - \hat{\partial}_i \hat{c}\right), \quad (3)$$

and $\hat{u}_i^P = \frac{D}{k_b T}\hat{\partial}_i \hat{\phi}$ is the phoretic contribution to the velocity [Kirby, 2010], where we assume the spatial dependency of the potential $\hat{\phi}$. To close the problem, the following boundary conditions are imposed on the membrane walls $\partial \mathbb{M}$:

$$\hat{u}_i = 0, \quad (4)$$

$$\hat{F}_i n_i^{\mathbb{M}} = \hat{S}_i n_i^{\mathbb{M}} \text{ on } \partial \mathbb{M}, \quad (5)$$

where $S_i$ is a known source term describing the solute interactions with the solid walls, possibly spatially varying along the boundary $\partial \mathbb{M}$ itself.

The interaction potential $\phi$ describes the interaction between surface, solvent and solute occurring at the molecular scale close to the solid boundaries of the membrane. Its range of action $\lambda$ typically spans distances from 0.5 to 200 nm from the solid surface, see Wood et al. [2004] and Kirby [2010] for further detail. In the short range potential case ($\lambda \ll \ell$), Michelin and Lauga [2014] embedded the effects of $\phi$ on the fluid flow in a boundary condition for the velocity field on $\partial \mathbb{M}$. In the present work, we first consider the full equations (i.e. long range potential case, $\lambda \approx \ell$), while the short range potential equations ($\lambda \ll \ell$) are studied in Section 4.

## 3 Homogenization of the long range potential equations

### 3.1 Non-dimensionalization of the flow equations

We first analyze the problem in the vicinity of the membrane (*inner* problem), defined in a microscopic elementary cell sketched in the inset of figure 1*a*. The variables in the microscopic domain are denoted by the superscript $\mathbb{I}$. The following non-dimensionalization is employed:

$$\hat{\tau} = \frac{\ell}{U^{\mathbb{I}}}\tau, \ \hat{c} = \Delta C^{\mathbb{I}} c^{\mathbb{I}}, \ \hat{x} = \ell x,$$

$$\hat{p} = \Delta P^{\mathbb{I}} p^{\mathbb{I}} = k_b T \Delta C^{\mathbb{I}} p^{\mathbb{I}}, \hat{u} = U^{\mathbb{I}} u^{\mathbb{I}} = \frac{\ell k_b T \Delta C^{\mathbb{I}}}{\mu} u^{\mathbb{I}}. \tag{6}$$

Note that the pressure scale $\Delta P^{\mathbb{I}}$ identifies the pressure difference caused by the solute concentration gradient. The dimensionless equations in the vicinity of the membrane thus read:

$$\begin{cases} Re_\ell(\partial_\tau u_i^{\mathbb{I}} + u_j^{\mathbb{I}} \partial_j u_i^{\mathbb{I}}) = -\partial_i p^{\mathbb{I}} + \partial_{ll}^2 u_i^{\mathbb{I}} - c^{\mathbb{I}} \partial_i \phi \\ \partial_i u_i^{\mathbb{I}} = 0 \\ Pe_\ell \partial_\tau c^{\mathbb{I}} + \partial_i F_i^{\mathbb{I}} = 0 \\ F_i^{\mathbb{I}} = Pe_\ell u_i^{\mathbb{I}} c^{\mathbb{I}} - (\partial_i c^{\mathbb{I}} + c^{\mathbb{I}} \partial_i \phi) \\ u_i^{\mathbb{I}} = 0 \text{ on } \partial \mathbb{M} \\ F_i n_i^{\mathbb{M}} = S_i n_i^{\mathbb{M}} \text{ on } \partial \mathbb{M}. \end{cases} \tag{7}$$

where $Re_\ell = \frac{\rho U^{\mathbb{I}} \ell}{\mu}$ and $Pe_\ell = \frac{U^{\mathbb{I}} \ell}{D}$ are the Reynolds and the Péclet number, both referred to the microscopic characteristic length $\ell$. Following the procedure sketched in figure 1*b*, the separation of scales allows us to introduce a fast (microscopic) and slow (macroscopic) variable through the decomposition $x_i \to x_i + \epsilon X_i$. Accordingly, spatial derivatives are transformed following the chain rule $\partial_i \to \partial_i + \epsilon \partial_I$, where the subscripts $i$ and $I$ respectively denote derivation with respect to $x_i$ and $X_i$. The unknown variables are expanded in series:

$$f^{\mathbb{I}} = \sum_{n=0}^{+\infty} \epsilon^n f^{\mathbb{I},n}(\mathbf{x}, \mathbf{X}, \tau), \tag{8}$$

where $f^{\mathbb{I}}$ represents the unknown velocity, pressure and concentration fields and $f^{\mathbb{I},n}$ denotes their *n*-th order approximation in the series expansion. Substituting the space variable decomposition and the form (8) into equation (7), and collecting each order term in $\epsilon$, at leading order, one obtains:

$$\begin{cases} -\partial_i p^{\mathbb{I},0} + \partial_{ll}^2 u_i^{\mathbb{I},0} - c^{\mathbb{I},0} \partial_i \phi = 0 \\ \partial_i u_i^{\mathbb{I},0} = 0 \\ \partial_i F_i^{\mathbb{I},0} = 0 \\ F_i^{\mathbb{I},0} = \partial_i c^{\mathbb{I},0} + c^{\mathbb{I},0} \partial_i \phi \\ u_i^{\mathbb{I},0} = 0 \text{ on } \partial \mathbb{M} \\ F_i^{\mathbb{I},0} n_i^{\mathbb{M}} = S_i n_i^{\mathbb{M}} \text{ on } \partial \mathbb{M}. \end{cases} \tag{9}$$

where we assumed $Re_\ell = Pe_\ell = O(\epsilon)$ and $Pe_\ell = O(\epsilon)$, i.e. the flow inertia and advective transport is negligible. This assumption is reasonable in a large range of biological flows [Jensen et al., 2016, Dvoriashyna et al., 2018] and engineering applications [Lee et al., 2014].

Equations are solved within the microscopic domain sketched in figure 1*a*, whose conditions on the boundaries $\mathbb{U}$ and $\mathbb{D}$ are the continuity of (*i*) solvent velocity and tractions, and (*ii*) solute concentration and flux:

$$\hat{u}_i^{in} = \hat{u}_i^{out}, \quad \hat{\Sigma}_{jk}^{in} n_k = \hat{\Sigma}_{jk}^{out} n_k, \quad \hat{c}^{in} = \hat{c}^{out}, \quad \hat{F}_i^{in} n_i = \hat{F}_i^{out} n_i, \tag{10}$$

with periodicity on the lateral sides of the unit cell. The superscript *out* denotes the outer region, while *in* denotes the inner region (cf. figure 1*a*).

For the outer problem, valid far from the membrane where the interaction potential effects can be neglected, we employ the following non-dimensionalization of the Navier-Stokes equations:

$$\hat{\tau} = \frac{L}{U^\mathbb{O}}\tau^\mathbb{O}, \quad \hat{c} = \Delta C^\mathbb{O} c^\mathbb{O}, \quad \hat{x} = Lx^\mathbb{O}, \quad \hat{p} = \Delta P^\mathbb{O} p^\mathbb{O}, \quad \hat{u}_i = U^\mathbb{O} u_i^\mathbb{O} = \frac{L\Delta P}{\mu} u_i^\mathbb{O}. \tag{11}$$

Once nondimensionalized, the boundary conditions linking the microscopic and macroscopic fields on $\mathbb{U}$ and $\mathbb{D}$ read:

$$u_i^{\mathbb{I},0} = u_i^{\mathbb{O},\mathbb{U}}, \quad \Sigma_{jk}^{\mathbb{I},0} n_k = \Sigma_{jk}^{\mathbb{O},\mathbb{U}} n_k, \quad F_i^{\mathbb{I},0} n_i = F_i^{\mathbb{O},\mathbb{U}} n_i, \quad \text{on } \mathbb{U}, \tag{12}$$

$$u_i^{\mathbb{I},0} = u_i^{\mathbb{O},\mathbb{D}}, \quad \Sigma_{jk}^{\mathbb{I},0} n_k = \Sigma_{jk}^{\mathbb{O},\mathbb{D}} n_k, \quad F_i^{\mathbb{I},0} n_i = F_i^{\mathbb{O},\mathbb{D}} n_i, \quad \text{on } \mathbb{D}. \tag{13}$$

In equations (12,13) the superscripts $\mathbb{O},\mathbb{U}$ or $\mathbb{O},\mathbb{D}$ denote outer quantities evaluated on $\mathbb{U}$ or $\mathbb{D}$, respectively. Conditions (12,13) close the linear problem (9), whose formal solution is presented in the next paragraph.

### 3.1.1 Microscopic solution

Since the problem is linear, we formally express the microscopic solution for the leading order flow approximation. We exploit the superposition principle for each source term in the governing equations [Zampogna and Gallaire, 2020, Zampogna et al., 2022, 2023]:

$$u_i^{\mathbb{I},0} = M_{ij} \Sigma_{jk}^{\mathbb{O},\mathbb{U}} n_k + N_{ij} \Sigma_{jk}^{\mathbb{O},\mathbb{D}} n_k + A_i F_k^{\mathbb{O},\mathbb{U}} n_k + B_i F_k^{\mathbb{O},\mathbb{D}} n_k + \alpha_i(\boldsymbol{S}), \tag{14}$$

$$p^{\mathbb{I},0} = Q_j \Sigma_{jk}^{\mathbb{O},\mathbb{U}} n_k + R_j \Sigma_{jk}^{\mathbb{O},\mathbb{D}} n_k + C F_k^{\mathbb{O},\mathbb{U}} n_k + D F_k^{\mathbb{O},\mathbb{D}} n_k + \beta(\boldsymbol{S}), \tag{15}$$

$$c^{\mathbb{I},0} = T F_k^{\mathbb{O},\mathbb{U}} n_k + Y F_k^{\mathbb{O},\mathbb{D}} n_k + \gamma(\boldsymbol{S}), \tag{16}$$

where $M_{ij}, N_{ij}, A_j, B_j, \alpha_i, Q_j, R_j, C, D, \beta, T, Y$, and $\gamma$ are yet unknown quantities stemming from the linearity of the solution. Notice that $\alpha_i, \beta$ and $\gamma$ linearly depend on the boundary source term **S**. We substitute equations (14-16) into the leading order problem (9) and boundary condition (12), we group terms by common forcings and derive several sets of equations to be satisfied independently of the forcings values. Therefore, we obtain several problems for $M_{ij}, N_{ij}, A_j, B_j, \alpha_i, Q_j, R_j, C, D, \beta, T, Y$, and $\gamma$ within the microscopic domain, presented now in detail. The problems below are written in the frame of reference of the membrane $(x_s, x_t, x_n)$.

Tensors $M_{ij}$ and $N_{ij}$ and vectors $Q_j$ and $R_j$ stem from the pure hydrodynamic problem and satisfy two independent microscopic problems presenting the same structure of Stokes equations, forced by unitary stress at the far-field boundaries, i.e.

$$\begin{cases} -\partial_i Q_j + \partial_{ll}^2 M_{ij} = 0 \text{ in } \mathbb{F} \\ \partial_i M_{ij} = 0 \text{ in } \mathbb{F} \\ M_{ij} = 0 \text{ on } \partial \mathbb{M} \\ \Sigma_{pq}(M_{\cdot j}, Q_j) n_q = \delta_{jp} \text{ on } \mathbb{U} \\ \Sigma_{pq}(M_{\cdot j}, Q_j) n_q = 0 \text{ on } \mathbb{D} \\ M_{ij}, Q_j \text{ periodic along } \boldsymbol{t} \text{ and } \boldsymbol{s} \end{cases} \quad \begin{cases} -\partial_i R_j + \partial_{ll}^2 N_{ij} = 0 \text{ in } \mathbb{F} \\ \partial_i N_{ij} = 0 \text{ in } \mathbb{F} \\ N_{ijk} = 0 \text{ on } \partial \mathbb{M} \\ \Sigma_{pq}(N_{\cdot j}, R_j) n_q = 0 \text{ on } \mathbb{U} \\ \Sigma_{pq}(N_{\cdot j}, R_j) n_q = \delta_{jp} \text{ on } \mathbb{D} \\ N_{ij}, R_j \text{ periodic along } \boldsymbol{t} \text{ and } \boldsymbol{s} \end{cases} \tag{17}$$

Tensors $M_{ij}$ and $N_{ij}$ represent the tensors which relate the velocity through the membrane with the macroscopic stresses acting on each side of the membrane [Zampogna and Gallaire, 2020]. Similarly, vectors $Q_j$ and $R_j$ relate the pressure field to the macroscopic stresses at each side of the membrane.

Scalars $T$ and $Y$ satisfy two independent Smoluchowski microscopic problems describing the passive transport of the solute concentration, i.e.

$$\begin{cases} \partial_{ii}^2 T + \partial_i \phi \partial_i T + T \partial_{ii}^2 \phi = 0 \text{ in } \mathbb{F} \\ (\partial_i T + T \partial_i \phi) n_i^\mathbb{M} = 0 \text{ on } \partial \mathbb{M} \\ -\partial_i T n_i = 1 \text{ on } \mathbb{U} \\ -\partial_i T n_i = 0 \text{ on } \mathbb{D} \\ T \text{ periodic along } \boldsymbol{t} \text{ and } \boldsymbol{s} \end{cases} \quad \begin{cases} \partial_{ii}^2 Y + \partial_i \phi \partial_i Y + Y \partial_{ii}^2 \phi = 0 \text{ in } \mathbb{F} \\ (\partial_i Y + Y \partial_i \phi) n_i^\mathbb{M} = 0 \text{ on } \partial \mathbb{M} \\ -\partial_i Y n_i = 0 \text{ on } \mathbb{U} \\ -\partial_i Y n_i = 1 \text{ on } \mathbb{D} \\ Y \text{ periodic along } \boldsymbol{t} \text{ and } \boldsymbol{s}, \end{cases} \tag{18}$$

where $n_i^\mathbb{M}$ denotes the normal to $\partial \mathbb{M}$ pointing towards the fluid domain. Problems (18) relate the concentration field close to the membrane to the concentration fluxes in the normal membrane direction.

Other microscopic problems emerge from the coupling between the velocity and concentration fields through the interaction potential

$$\begin{cases} -\partial_i C + \partial_{ll}^2 A_i + T\partial_i \phi = 0 \text{ in } \mathbb{F} \\ \partial_i A_i = 0 \text{ in } \mathbb{F} \\ A_i = 0 \text{ on } \partial\mathbb{M} \\ \Sigma_{pq}(A., C)n_q = 0 \text{ on } \mathbb{U} \text{ and } \mathbb{D} \\ A_i, C \text{ periodic along } t \text{ and } s \end{cases} \quad \begin{cases} -\partial_i D + \partial_{ll}^2 B_i + Y\partial_i \phi = 0 \text{ in } \mathbb{F} \\ \partial_i B_i = 0 \text{ in } \mathbb{F} \\ B_i = 0 \text{ on } \partial\mathbb{M} \\ \Sigma_{pq}(B., D)n_q = 0 \text{ on } \mathbb{U} \text{ and } \mathbb{D} \\ B_i, D \text{ periodic along } t \text{ and } s. \end{cases} \quad (19)$$

$A_i$, $B_i$, $C$, $D$ quantify the velocity and pressure contributions induced by the concentration gradient within the fluid domain close to the membrane. The associated problems are formally analogous to the Stokes equations with homogeneous boundary conditions but require the solution of (18) beforehand. Indeed, the concentration-related volumetric terms in the momentum equations ($T\partial_i\phi$ and $Y\partial_i\phi$) act as source terms.

The phoretic boundary contributions solve the following coupled problems

$$\begin{cases} \partial_{ii}^2 \gamma + \partial_i\phi\partial_i\gamma + \gamma\partial_{ii}^2\phi = 0 \text{ in } \mathbb{F} \\ (\partial_i\gamma + \gamma\partial_i\phi)n_i^{\mathbb{M}} = S_i n_i^{\mathbb{M}} \text{ on } \partial\mathbb{M} \\ -\partial_i\gamma n_i = 0 \text{ on } \mathbb{U} \text{ and } \mathbb{D} \\ \gamma \text{ periodic along } t \text{ and } s \end{cases} \quad \begin{cases} -\partial_i\beta + \partial_{ll}^2\alpha_i + \gamma\partial_i\phi = 0 \text{ in } \mathbb{F} \\ \partial_i\alpha_i = 0 \text{ in } \mathbb{F} \\ \alpha_i = 0 \text{ on } \partial\mathbb{M} \\ \Sigma_{pq}(\alpha., \beta)n_q = 0 \text{ on } \mathbb{U} \text{ and } \mathbb{D} \\ \alpha_i, \beta \text{ periodic along } t \text{ and } s, \end{cases} \quad (20)$$

where $\beta$ plays a role analogous to the pressure in the Stokes-like problem for $\alpha_i$. The boundary condition on $\partial\mathbb{M}$ in the problem for $\gamma$ contains the source term $S_i = S_i(x)$. Quantity $\gamma$ is the phoretic contribution to $c^{\mathbb{I},0}$ and acts as a source term for the concentration field. Vector $\alpha_i$ instead represents the phoretic membrane velocity contribution purely induced by concentration differences on $\partial\mathbb{M}$ due to the solute-membrane interactions embedded in $\gamma$.

Equations (14-16) with the solvability conditions (17-20) forms a closed physico-chemical description of the transport across the membrane. However, the formal solution (14-16) to (9) depends on the fast and slow variables, while our final objective consists of developing purely macroscopic interface conditions valid on the homogeneous membrane $\mathbb{C}$ (cf. figure 1c). In the next paragraph, we therefore introduce an averaging step.

### 3.1.2 From the microscopic solution to the macroscopic interface condition

To filter the microscale dependence in equations (14-16), the upward $\overline{\cdot}^-$ and downward $\overline{\cdot}^+$ (surface) averages are introduced [Zampogna et al., 2023]:

$$\overline{\cdot}^- = \lim_{x_n \to +\infty} \frac{1}{|\mathbb{U}|} \int_{\mathbb{U}} \cdot \, dx_s dx_t \quad \text{and} \quad \overline{\cdot}^+ = \lim_{x_n \to -\infty} \frac{1}{|\mathbb{D}|} \int_{\mathbb{D}} \cdot \, dx_s dx_t. \quad (21)$$

The introduction of upward and downward averages allows for discontinuous macroscopic velocity and concentration profiles across the membrane, crucial in the problem analyzed here since the main fluid flow driving force is represented by microscopic variations in the solute concentration field. By applying averages (21) to (14-16), we obtain the following set of equations for the velocity and concentration field at the upward ($\mathbb{C}^-$) and downward ($\mathbb{C}^+$) sides of the membrane:

$$u_i^- := \overline{u^{\mathbb{I},0}}_i^- = \mathcal{M}_{ij}^- \Sigma_{jk}^{\mathbb{C}^-} n_k + \mathcal{N}_{ij}^- \Sigma_{jk}^{\mathbb{C}^+} n_k + \mathcal{A}_i^- F_k^{\mathbb{C}^-} n_k + \mathcal{B}_i^- F_k^{\mathbb{C}^+} n_k + \alpha_i^-(S), \quad (22)$$

$$u_i^+ := \overline{u^{\mathbb{I},0}}_i^+ = \mathcal{M}_{ij}^+ \Sigma_{jk}^{\mathbb{C}^-} n_k + \mathcal{N}_{ij}^+ \Sigma_{jk}^{\mathbb{C}^+} n_k + \mathcal{A}_i^+ F_k^{\mathbb{C}^-} n_k + \mathcal{B}_i^+ F_k^{\mathbb{C}^+} n_k + \alpha_i^+(S), \quad (23)$$

$$p^- := \overline{p^{\mathbb{I},0}}^- = Q_j^- \Sigma_{jk}^{\mathbb{C}^-} n_k + \mathcal{R}_j^- \Sigma_{jk}^{\mathbb{C}^+} n_k + C^- F_k^{\mathbb{C}^-} n_k + \mathcal{D}^- F_k^{\mathbb{C}^+} n_k + \beta^-(S), \quad (24)$$

$$p^+ := \overline{p^{\mathbb{I},0}}^+ = Q_j^+ \Sigma_{jk}^{\mathbb{C}^-} n_k + \mathcal{R}_j^+ \Sigma_{jk}^{\mathbb{C}^+} n_k + C^+ F_k^{\mathbb{C}^-} n_k + \mathcal{D}^+ F_k^{\mathbb{C}^+} n_k + \beta^+(S), \quad (25)$$

$$c^- := \overline{c^{\mathbb{I},0}}^- = \mathcal{T}^- F_k^{\mathbb{C}^-} n_k + \mathcal{Y}^- F_k^{\mathbb{C}^+} n_k + \gamma^-(S), \quad (26)$$

$$c^+ := \overline{c^{\mathbb{I},0}}^+ = \mathcal{T}^+ F_k^{\mathbb{C}^-} n_k + \mathcal{Y}^+ F_k^{\mathbb{C}^+} n_k + \gamma^+(S), \quad (27)$$

where

$$\mathcal{M}_{ij}^- = \overline{M}_{ij}^- - x_n|_{\mathbb{U}}(\delta_{it}\delta_{jt} + \delta_{is}\delta_{js}), \qquad \mathcal{M}_{ij}^+ = \overline{M}_{ij}^+, \qquad Q_{ij}^\pm = \overline{Q}_{ij}^\pm,$$

$$\mathcal{A}_i^\pm = \overline{A}_i^\pm, \qquad \mathcal{B}_i^\pm = \overline{B}_i^\pm, \qquad C^\pm = \overline{C}^\pm, \qquad \mathcal{D}^\pm = \overline{D}^\pm,$$

| | $\mathcal{M}_{tt}$ | $\mathcal{M}_{nn}$ | $\mathcal{N}_{tt}$ | $\mathcal{N}_{nn}$ | $\mathcal{T}$ | $\mathcal{Y}$ |
|---|---|---|---|---|---|---|
| $\mathbb{U}$ | $4.76 \times 10^{-2}$ | $4.98 \times 10^{-3}$ | $4.1 \times 10^{-6}$ | $-4.98 \times 10^{-3}$ | $1.20 \times 10^{-2}$ | $-1.07 \times 10^{-3}$ |
| $\mathbb{D}$ | $4.1 \times 10^{-6}$ | $4.98 \times 10^{-3}$ | $4.76 \times 10^{-2}$ | $-4.98 \times 10^{-3}$ | $1.07 \times 10^{-3}$ | $-1.20 \times 10^{-2}$ |

Table 1: Non-zero averaged components of $\mathcal{M}, \mathcal{N}, \mathcal{T}$ and $\mathcal{Y}$

$$\begin{aligned}
\mathcal{T}^- &= \overline{T}^- - x_n|_{\mathbb{U}}, \qquad \mathcal{T}^+ = \overline{T}^+ \\
\mathcal{N}_{ij}^- &= \overline{N}_{ij}^-, \qquad \mathcal{N}_{ij}^+ = \overline{N}_{ij}^+ - x_n|_{\mathbb{D}}(\delta_{it}\delta_{jt} + \delta_{is}\delta_{js}), \qquad \mathcal{R}_{ij}^\pm = \overline{R}_{ij}^\pm, \\
\mathcal{Y}^- &= \overline{Y}^-, \qquad \mathcal{Y}^+ = \overline{Y}^+ - x_n|_{\mathbb{D}}. \\
\alpha_i^\pm &= \overline{\alpha}_i^\pm, \qquad \beta^\pm = \overline{\beta}^\pm, \qquad \gamma^\pm = \overline{\gamma}^\pm,
\end{aligned} \qquad (28)$$

The velocity field at the membrane is given by several contributions. If only the first two terms on the RHS in equations (22,23) are considered, the problem reduces to a pure hydrodynamic problem governed by the Stokes equations across the membrane [Zampogna et al., 2023]. The condition states that the macroscopic velocity at one side of the membrane is proportional to the stresses at both sides of the membrane. The velocity in the normal membrane direction is reminiscent of the Darcy law, while the tangential velocity is governed by a Navier slip condition [Ledda et al., 2021]. Therefore, these proportionality coefficients, $\mathcal{M}_{ij}^\pm$ and $\mathcal{N}_{ij}^\pm$, are also called the upward and downward *Navier* tensors, where both permeability ($\mathcal{M}_{nn}^\pm$ and $\mathcal{N}_{nn}^\pm$) and slip ($\mathcal{M}_{tt}^\pm, \mathcal{M}_{ss}^\pm, \mathcal{N}_{tt}^\pm$ and $\mathcal{N}_{ss}^\pm$) terms appear.

Analogously, if we consider only the passive transport of a chemical species, equations (26,27) describe the concentration jump at the membrane. The scalars $\mathcal{T}^\pm$ and $\mathcal{Y}^\pm$ represent the *effective diffusivity* of the solute in the normal membrane direction, deviating from the bulk one due to the membrane microstructure.

The coupling given by the interaction potential gives rise to several additional terms. The velocity field depends also on the concentration flux across the membrane due to the presence of the potential $\phi$ in the leading order problem (9): vectors $\mathcal{A}_i^\pm$ and $\mathcal{B}_i^\pm$ indeed represent the solute gradient effect on the solvent flow and can be defined as the *bulk fluid mobility* across the membrane. Eventually, term $\alpha_i^\pm$ represents the *surface phoretic* velocity generated by the solute concentration variation at the solid walls of the membrane inclusions, and $\gamma$ is an additional concentration stemming from the solute source at the membrane walls.

## 3.2 Solution and validation

In this section, we compare our homogenized interface condition against full-scale cases. We consider a 2D phoretic membrane composed of an array of circular inclusions. The membrane splits into two parts a two-dimensional U-shaped channel and is located at its center. The membrane releases solute particles at an imposed rate $\mathbf{S}(\mathbf{x})$ (cf. problem (20)), inducing a concentration gradient that drives the flow within the channel. A thorough review of typical potential functions for specific colloid-surface interactions can be found in Elimelech et al. [1995]. Following Gebäck and Heintz [2019], we employ the potential $\phi$ defined as

$$\phi(\mathbf{x}) = \frac{\mathfrak{A}}{2}\left(1 - \tanh\left(\frac{d(\mathbf{x}) - a}{\delta'}\right)\right) \qquad (29)$$

where $d(x)$ is the distance function from the center of the solid inclusion of radius $a$. In our case $\mathfrak{A} = 10$ and $\delta' = \lambda/(1-a) = 0.0025$. These parameters' values fall in the range considered by Gebäck and Heintz [2019]. The source term $\mathbf{S}$ in the solute flux specified in equation (5) is parameterized using an angle $\theta$ which expresses a parametric rotation of the source $S_i$ to the axis $x_n$ as follows:

$$\mathbf{S}(\theta) = C_0 + cos(\theta)\mathbf{e}_n + sin(\theta)\mathbf{e}_t, \qquad (30)$$

where $C_0 > 0$ is a constant value that allows us to avoid non-physical negative concentrations. For example, when $\theta = 0$, then $S_i$ is directed along $x_n$, while when $\theta = \pi/2$, it is directed along $x_t$. This simple form is employed to appreciate the effects of the source orientation on the microscopic and macroscopic flow.

### 3.2.1 Microscopic solution

We present the solution of problems (17-20) obtained in section 3.1. We initially focus on the case of a solute release along the normal membrane direction, i.e. $\theta = 0$ in equation (30). Additionally, in the same equation, we specify $C_0 = 5$ in $\Delta C^{\mathbb{I}}/\ell$ units. This assumption allows us to avoid unphysical negative values of solute concentration caused by a potentially negative value of some entries of **S**. As stated, these problems are defined within the microscopic elementary cell introduced in figure 1a. The microscopic spatial variables are denoted by $(x_t, x_n)$.

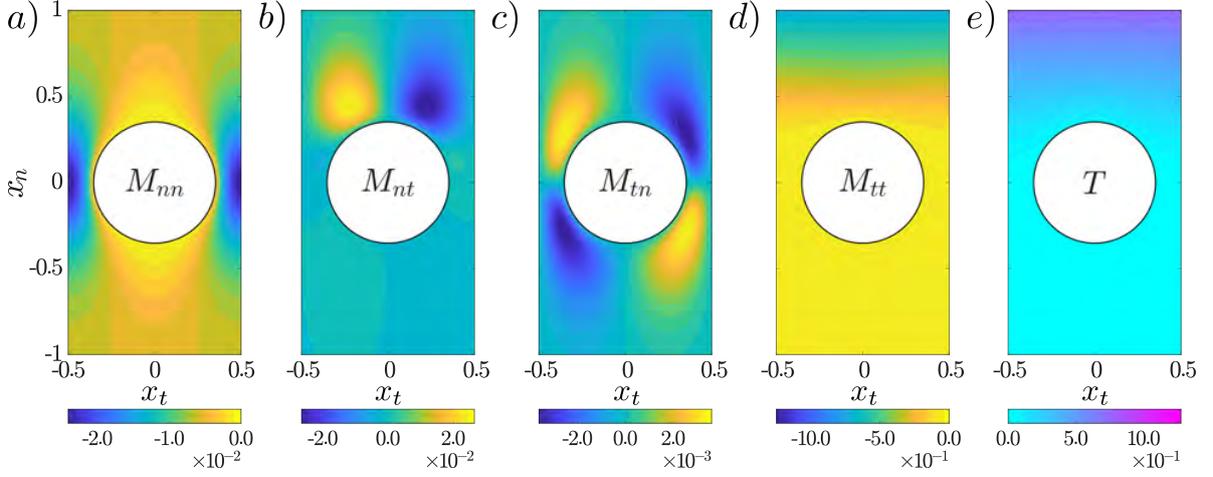

Figure 2: Microscopic fluid flow-related $M_{ij}$ and concentration-related $T$ fields arising from the solvent-to-solute interactions. These fields are not influenced by the source term **S**. Since $N_{ij}(x_i) = -M_{ij}(-x_i)$ and $Y(x_i) = -T(-x_i)$ for symmetric geometries, these fields are not shown.

|  | $\gamma$ | $\mathcal{A}_n$ | $\mathcal{B}_n$ | $\alpha_n$ |
|---|---|---|---|---|
| $\mathbb{U}$ | $2.13 \times 10^{-1}$ | $1.71 \times 10^{-7}$ | $-1.71 \times 10^{-7}$ | $-2.14 \times 10^{-7}$ |
| $\mathbb{D}$ | $1.58 \times 10^{-1}$ | $1.71 \times 10^{-7}$ | $-1.71 \times 10^{-7}$ | $-2.14 \times 10^{-7}$ |

Table 2: Non-zero averaged components of $\gamma$, $A$, $B$ and $\alpha$ for $\theta = 0$ in the source term of the microscopic problem (20) as defined in (30)

Figure 2 depicts the isocontours of $M_{ij}$ and $T$. Because of the symmetry of the microscopic inclusion, $N_{ij}$ and $Y$ are not shown since $N_{ij}(x_k) = -M_{ij}(-x_k)$ and $Y(x_k) = -T(-x_k)$. In addition, the off-diagonal components of the upward Navier slip tensor $M_{ij}$ are anti-symmetric (figure 2b, c). Components $M_{nn}$ and $M_{tn}$ resemble the Stokes flow velocity past a cylinder when the flow is driven by a pressure gradient. conversely, $M_{tt}$ and $M_{nt}$ suggest the development of a linear profile of velocity due to tangential stress at the top. Iso-levels of $T$ are shown in figure 2e. Despite the local effect of the potential (see inset), there is a difference of one order of magnitude in $T$ between the two sides of the solid inclusion.

Quantities $A_i$, $\alpha_i$ and $\gamma$ are shown in figure 3. Vector $B_i$ is not represented since $B_i(x_k) = -A_i(-x_k)$, for this symmetric configuration. A closer look into the microscopic $A_n$ field (figure 3a) unveils the presence of the so-called *plug flow* velocity structure in the narrow region between two inclusions, typical of flows induced by osmotic or phoretic gradients where the fluid forcing has its maximum close to the walls Kirby [2010]. In the present case, this flow structure is associated with the gradient of $T$ (which, in turn, can be linked to the microscopic solute transport). The tangential contribution $A_t$ (figure 3b) is instead anti-symmetric due to the symmetry of the membrane solid inclusion.

The normal-to-the-membrane component of $\alpha$ (figure 3c), exhibits a plug flow velocity profile. In analogy with **A**, also the tangent component $\alpha_t$ is anti-symmetric (figure 3d). The plug flow distribution agrees with the interpretation of $\alpha$, the solvent velocity contribution due to solute gradients induced by the solid walls, modeled here with the introduction of the release term $S_i$ (cf. microscopic problem (20)). Eventually, the scalar field $\gamma$ (figure 3e) represents an additional concentration stemming from solid-fluid interactions at the boundaries, which here highlights the anti-symmetry of the imposed flux $S_i$ at the solid walls.

At this stage, the reader might wonder how $S_i$ influences the microscopic flow and concentration structures. An intuitive answer is provided in figure 4(*a* − *c*), where the microscopic quantities influenced by $S_i$ are reported, for a value of $\theta = \pi/4$. The release term $S_i$ influences $\gamma$, whose anti-symmetry axis is not aligned with $x_n$ as in figure 3e. Furthermore, flow structures associated with $\alpha$ are rotated, leading to a net tangential-to-the-membrane velocity contribution due to $\alpha_t$. All other microscopic quantities are not influenced by a rotation of **S** since they are not linked to the solute concentration gradients at the solid boundaries.

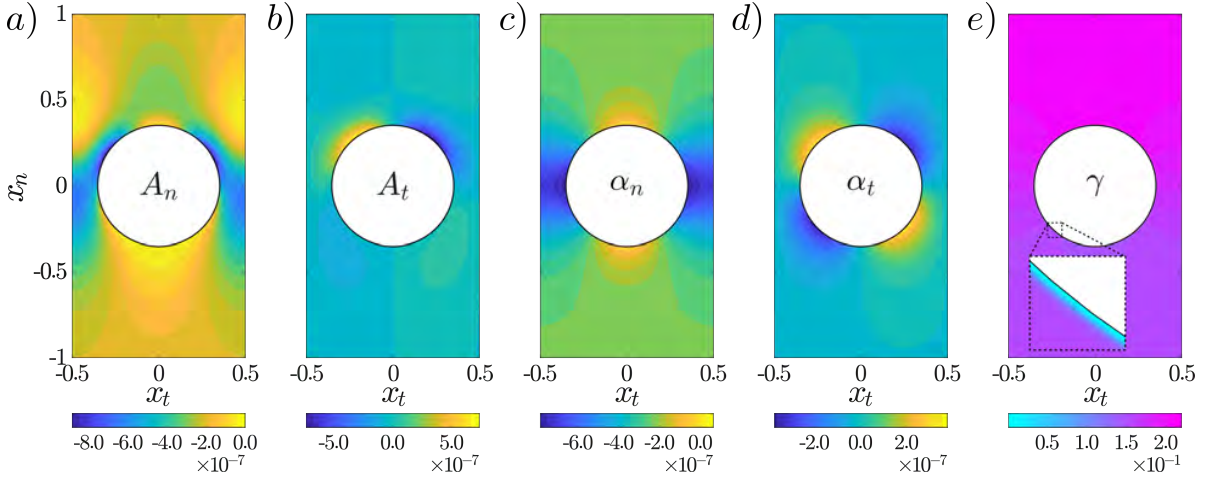

Figure 3: Microscopic fluid flow-related $A_i$, $\alpha_i$ and concentration-related $\gamma$ fields arising from the solvent-to-solute interactions. The $\alpha_i$, $\gamma$ fields are influenced by the source term **S**. $\theta = 0$ and $C_0 = 5$ in the present case. Since $A_i = -B_i$ for symmetric geometries, these fields are not shown.

|   | $\gamma$ | $\alpha_t$ | $\alpha_n$ |
|---|---|---|---|
| $\cdot^{\mathbb{U}}$ | $5.31 \times 10^{-2}$ | $-6.19 \times 10^{-5}$ | $-3.63 \times 10^{-5}$ |
| $\cdot^{\mathbb{D}}$ | $6.69 \times 10^{-2}$ | $-6.19 \times 10^{-5}$ | $-3.63 \times 10^{-5}$ |

Table 3: $\gamma$ and $\alpha_i$ with $S_i(\boldsymbol{x})$ defined in equation (30) for $\theta = \pi/4$ and used in microscopic problem (20).

The microscopic solution fields are exploited to obtain their upward and downward averages (21). The non-zero entries of these quantities are listed in tables 1, 2, and 3. As already noticed in Zampogna et al. [2023] the upward and downward averages of $M_{nn}$ ($N_{nn}$) are the same, due to the conservation of mass across the membrane, for normal stress forcing. This property is extended here also to $A_n$ ($B_n$) and $\alpha_n$ since they contribute to the macroscopic normal-to-the-membrane velocity that must be conserved since there is no net flow rate at the solid boundary. A cross-comparison between tables 2 and 3 shows how a non-symmetric $S_i$ produces a functional membrane anisotropy, quantified by a non-zero $\alpha_t$.

The effect of the concentration boundary forcing on $\alpha$ and $\gamma$ is investigated through a parametric study for $\theta$. The averaged values are sketched in figure 4(d,e) for $\theta \in [0, 2\pi]$. From an analysis of these curves, one could *a priori* guess the membrane flow behavior, independently of the macroscopic flow configuration. Assuming that the only driving force of the flow is the concentration gradient at the membrane walls (i.e., $S_i$), the dashed red lines suggest that, for $\theta = \pi/2$ and $\theta = 3\pi/2$, the normal-to-the-membrane solvent velocity is zero while the tangential one is maximum, thus denoting a pure tangential flow. *Vice versa*, for $\theta = 0$ and $\theta = 2\pi$, the solute gradient produces a pure normal-to-the-membrane flow.

These indications suggest that homogenization can provide macroscopic guidelines that can be included in a hypothetical design procedure. In the next section, we show how these microscopic averaged quantities influence the macroscopic behavior of the flow across the membrane.

### 3.2.2 Macroscopic solutions and comparison with full-scale simulations

We compare the full-scale solution of equations (7) with the macroscopic solution obtained by substituting, to the real membrane microstructure, a fictitious interface where conditions (22– 27) apply. We consider the flow within the U-shaped two-dimensional channel sketched in figure 5. The channel's upper left and upper right sides are open and the fluid is free to flow across them. The duct is split in the middle by a phoretic membrane, whose microstructure is formed by an array of cylinders with a fluid-to-total ratio on the membrane centerline of 0.3. The microscopic membrane properties are varied to assess the capability of the model to catch anisotropic membrane behaviors. In the first configuration (figure 5), the membrane is formed by 10 solid inclusions ($\epsilon = 0.1$) and $\theta = 0$ in the source term $S_i$ of equation (30). Figure 5 depicts the contours of the concentration field (magenta-to-cyan) for the full-scale (left) and macroscopic solution (right). In the same figure, the flow streamlines are colored by the velocity magnitude in gray scale. The qualitative good agreement evinced from figure 5 is quantified in figure 6, where we sample the solution of the upward and downward membrane sides, immediately outside of the potential range. The normal (tangential)-to-the membrane velocity exhibits a symmetric (anti-symmetric) behavior about the membrane centerline, while the concentration field presents a sharp jump generated by the source $S_i$. The macroscopic model (yellow and purple lines) reproduces well the averages (black symbols) of the full-scale solution (black lines). We notice only a discrepancy in the concentration values (frame $d$), however, the concentration jump is correctly captured by the macroscopic model with a relative error of 0.9%.

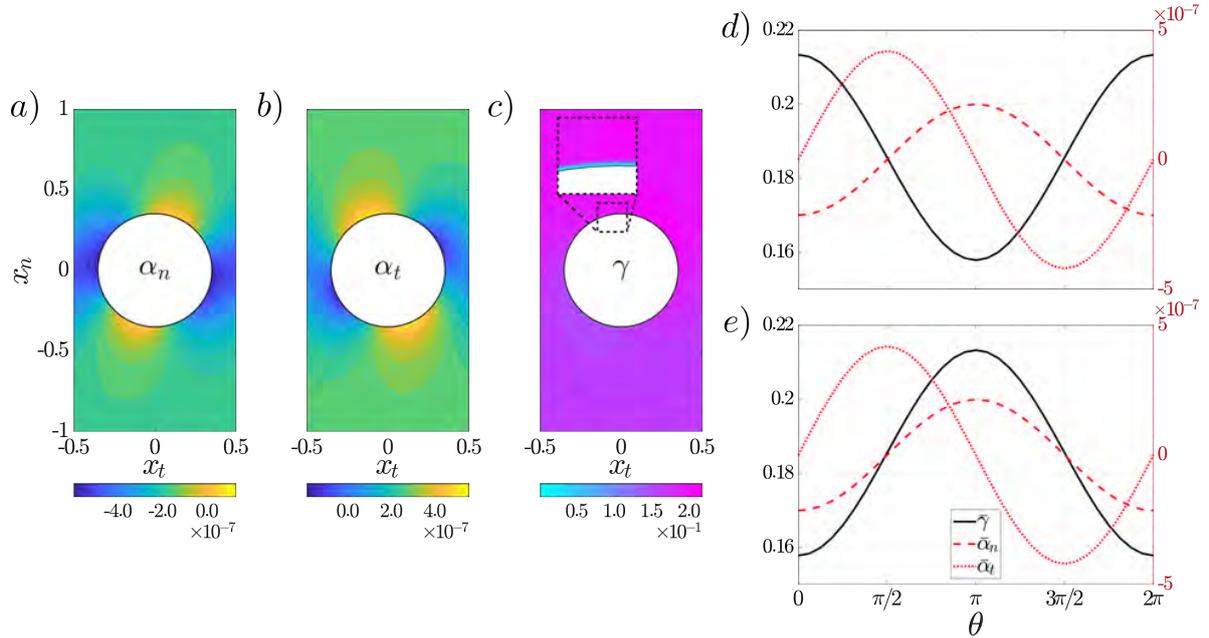

Figure 4: (a,b,c) Microscopic fluid flow-related $\alpha_i$ and concentration-related $\gamma$ fields arising from the solvent-to-solute interactions. These fields are influenced by the source term $S$. $\theta = \pi/4$ in the present case. (d,e) Average $\gamma$ and $\alpha_i$ values computed on the top ($a$) and bottom ($b$) sides of the solid inclusion as a function of the source rotation angle $\theta$, see equation (30).

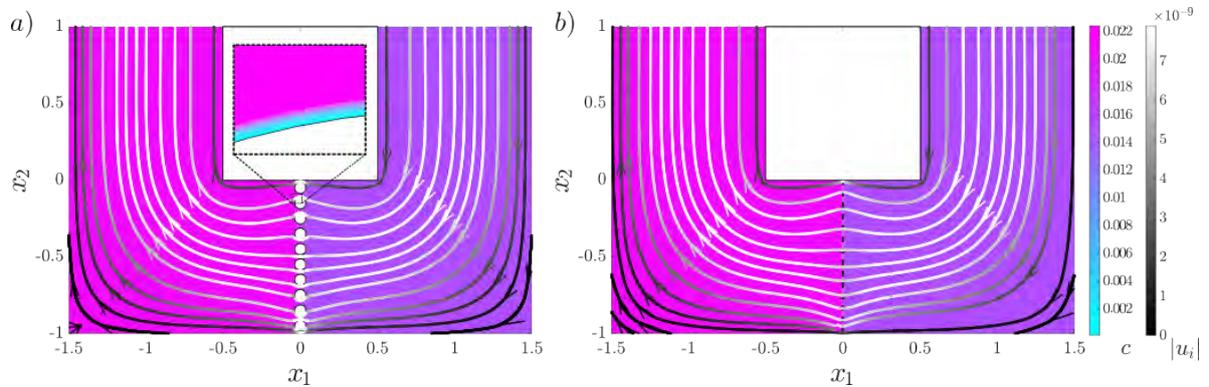

Figure 5: Comparison between the full-scale (panel $a$) and macroscopic ($b$) solutions in terms of solute concentration (iso-levels) and fluid flow velocity (streamlines colored by velocity magnitude) in the U-shaped channel when $\theta = 0$ and $\epsilon = 0.1$.

In a second configuration, we modify the source term $S$, imposing $\theta = \pi/4$, while maintaining the same geometrical properties of the microstructure. In this case, the concentration gradient is not fully normal to the membrane and a non-zero vertical velocity is generated (figure 7a,b). The presence of a vertical velocity produces two recirculation regions near the edges of the membrane. These vortices are smaller in the macroscopic simulation. Figure 7(c-f) shows that the flow in the close vicinity of the edges is not periodic. Therefore, while the macroscopic model well reproduces the flow behavior in the internal part of the membrane, the breakdown of the periodicity near the membrane edges produces slight macroscopic deviations. Concerning concentration fields, although the final concentrations are different, again the model correctly reproduces the concentration jump across the membrane with a relative error of about 3%. Despite these small deviations, likely mainly induced by the use of periodic microscopic problems (see e.g. Ledda et al., 2021), the model faithfully reproduces the macroscopic behavior of the flow, including hydrodynamics, transport and phoretic effects, only relying on interface conditions closed by known quantities.

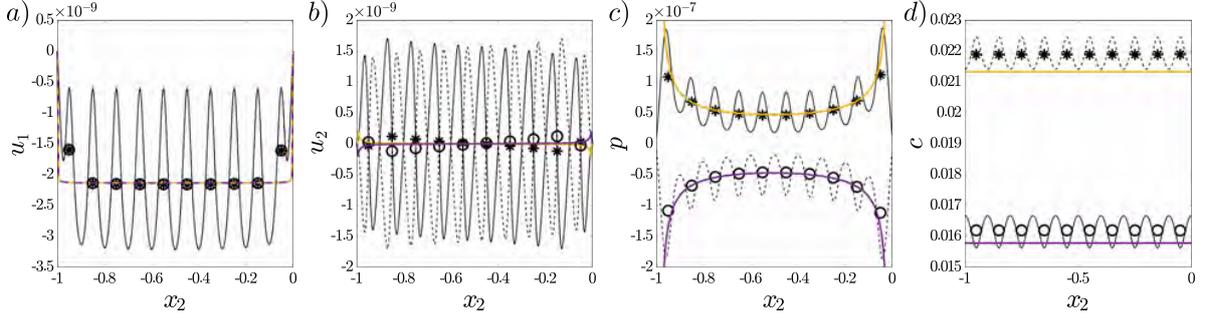

Figure 6: Comparison between the full-scale (point-wise fields: black solid and dashed line for the left and right-hand side, respectively; average fields: black asterisks and circles for the left and right-hand sides of the membrane, respectively) and macroscopic (yellow and purple for the left and right sides of the membrane, respectively) velocity components ($a, b$), pressure ($c$) and solute concentration ($d$) on the membrane sides when $\theta = 0$ and $\epsilon = 0.1$.

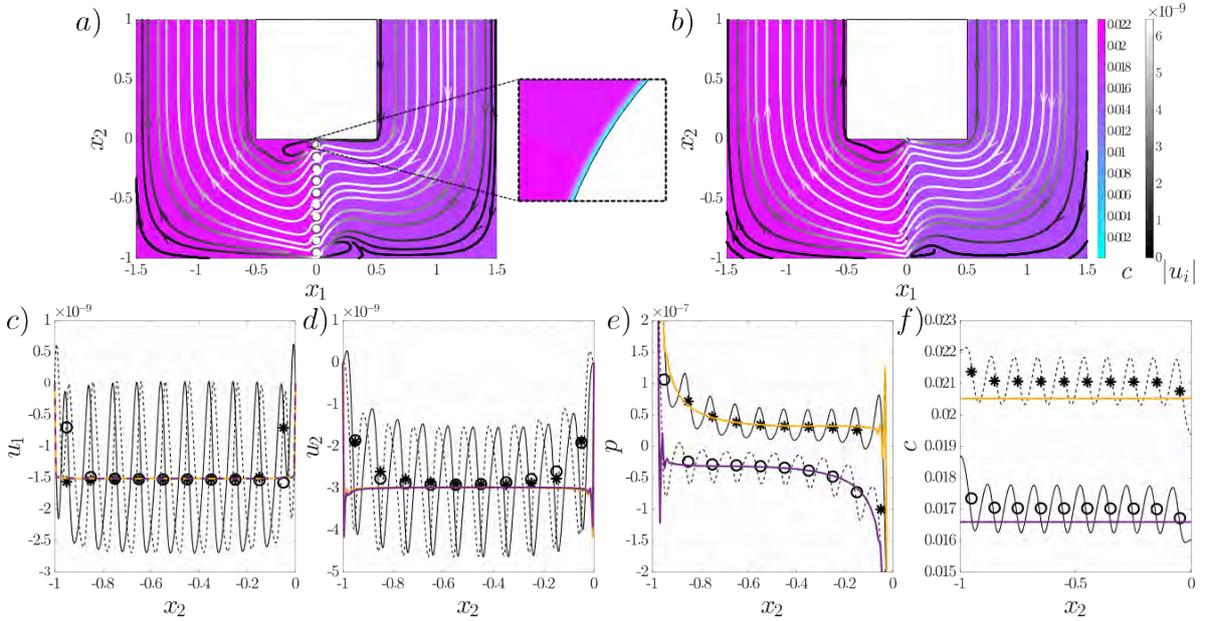

Figure 7: $\theta = \pi/4$ and $\epsilon = 0.1$. (a,b) Same as Figure 5. (c,d,e) Comparison between the full-scale averaged fields (black asterisks and circles for the left and right-hand sides of the membrane, respectively) and macroscopic (yellow and purple for the left- and right-hand sides of the membrane, respectively) velocity components and solute concentration on the membrane.

## 4 Short range potential limit

In the case of short-range potential, and $\lambda \ll \ell$, the effects of the interaction potential $\phi$ are felt only in the close vicinity of the solid boundary while they can be neglected in the bulk fluid. The governing equations degenerate in a coupled Stokes-Smoluchowski equation without the interaction potential, as in Zampogna et al. [2023]. However, a boundary layer analysis allows us to embed the interaction potential effects in a phoretic slip boundary condition for the solvent velocity and an effective solute flux at the membrane walls, see Michelin and Lauga [2014] for further detail. The full problem hence reads

$$\begin{cases} Re_\ell u_j^\mathbb{I} \partial_j u_i^\mathbb{I} = -\partial_i p^\mathbb{I} + \partial_{ll}^2 u_i^\mathbb{I} \\ \partial_i u_i^\mathbb{I} = 0 \\ \partial_i F_i^\mathbb{I} = 0 \\ F_i^\mathbb{I} = Pe_\ell u_i^\mathbb{I} c^\mathbb{I} - D \partial_i c^\mathbb{I} \\ u_i^\mathbb{I} = M(\delta_{ij} - n_i^\mathbb{M} n_j^\mathbb{M}) \partial_j c^\mathbb{I} \text{ on } \partial \mathbb{M} \\ c^\mathbb{I} = S_i(X) n_i^\mathbb{M} \text{ on } \partial \mathbb{M}, \end{cases} \quad (31)$$

where $M$ is the motility scalar as in Michelin and Lauga [2014]. The source condition on the solute at the membrane walls $\partial \mathbb{M}$ is specified here in terms of concentration and not in terms of fluxes. However, one could introduce supplementary parameters to specify the source as a concentration flux as already done in Zampogna et al. [2023]. Applying the same procedure as in §3 we obtain the following leading order problem

$$\begin{cases} -\partial_i p^{\mathbb{I},0} + \partial_{ll}^2 u_i^{\mathbb{I},0} = 0 \\ \partial_i u_i^{\mathbb{I},0} = 0 \\ \partial_i F_i^{\mathbb{I},0} = 0 \\ F_i^{\mathbb{I},0} = -\partial_i c^{\mathbb{I},0} \\ u_i^{\mathbb{I},0} = M(\delta_{ij} - n_i^\mathbb{M} n_j^\mathbb{M}) \partial_j c^{\mathbb{I},0} \text{ on } \partial \mathbb{M} \\ c^{\mathbb{I},0} = S_i(X) n_i^\mathbb{M} \text{ on } \partial \mathbb{M}, \end{cases} \quad (32)$$

since we assumed that $Re_\ell$ and $Pe_\ell$ are of order $O(\epsilon)$ Zampogna et al. [2022]. Following the same steps as in §3, problem (32) is closed by the far-field boundary conditions (12,13) on the sides $\mathbb{U}$ and $\mathbb{D}$ of the microscopic elementary cell, which are not affected by the presence of the interaction potential. Applying averages (21) to the formal solution of problem (32), adopting the notation introduced in equation (28) and neglecting the superscripts $\mathbb{I}, 0$, we obtain

$$u_i^- = \mathcal{M}_{ij}^- \Sigma_{jk}^{\mathbb{C}^-} n_k + \mathcal{N}_{ij}^- \Sigma_{jk}^{\mathbb{C}^+} n_k + \overline{u_i^{ph}}^- \quad (33)$$

$$u_i^+ = \mathcal{M}_{ij}^+ \Sigma_{jk}^{\mathbb{C}^-} n_k + \mathcal{N}_{ij}^+ \Sigma_{jk}^{\mathbb{C}^+} n_k + \overline{u_i^{ph}}^+ \quad (34)$$

$$p^- = Q_j^- \Sigma_{jk}^{\mathbb{C}^-} n_k + \mathcal{R}_j^- \Sigma_{jk}^{\mathbb{C}^+} n_k + \overline{p^{ph}}^- \quad (35)$$

$$p^+ = Q_j^+ \Sigma_{jk}^{\mathbb{C}^-} n_k + \mathcal{R}_j^+ \Sigma_{jk}^{\mathbb{C}^+} n_k + \overline{p^{ph}}^+ \quad (36)$$

$$c^- = \mathcal{T}^- F_k^{\mathbb{C}^-} n_k + \mathcal{Y}^- F_k^{\mathbb{C}^+} n_k + \overline{c^{ph}}^- \quad (37)$$

$$c^+ = \mathcal{T}^+ F_k^{\mathbb{C}^-} n_k + \mathcal{Y}^+ F_k^{\mathbb{C}^+} n_k + \overline{c^{ph}}^+ \quad (38)$$

where $u_i^{ph}$, $p^{ph}$ and $c^{ph}$, solve the following microscopic problems

$$\begin{cases} -\partial_i p^{ph} + \partial_{ll}^2 u_i^{ph} = 0 \text{ in } \mathbb{F} \\ \partial_i u_i^{ph} = 0 \text{ in } \mathbb{F} \\ u_i^{ph} = M(\delta_{ij} - n_i^\mathbb{M} n_j^\mathbb{M}) \cdot \partial_j c^\mathbb{M} \text{ on } \partial \mathbb{M} \\ \Sigma_{pq}(u_i^{ph}, p^{ph}) n_q = 0 \text{ on } \mathbb{U}, \mathbb{D} \\ u_i^{ph}, p^{ph} \text{ periodic along } t, s, \end{cases} \quad \begin{cases} \partial_{ii}^2 c^{ph} = 0 \text{ in } \mathbb{F} \\ \xi \partial_i c^{ph} n_i^\mathbb{M} + \psi c^{ph} = S_i(X) n_i^\mathbb{M} \text{ on } \partial \mathbb{M} \\ \partial_n c^{ph} = 0 \text{ on } \mathbb{U}, \mathbb{D} \\ c^{ph} \text{ periodic along } t, s, \end{cases} \quad (39)$$

while tensors $M_{ij}$ and $N_{ij}$ and scalars $T$ and $Y$ solve, respectively, problems (17) and (18), of the term involving $\phi$ disappeared and the boundary condition on $\partial \mathbb{M}$ reduces to $T = Y = 0$. In the short-range potential limit, new quantities $p^{ph}, u_i^{ph}$ and $c^{ph}$ are introduced in place of $A_i, B_i, C, D, \gamma, \alpha_i$ and $\eta$. The effect of the potential is now embedded in two non-homogeneous boundary conditions on $\partial \mathbb{M}$ in problems (39). As mentioned at the beginning of the section, the short and long-range descriptions are equivalent if $\lambda \ll L$. However, for a given source term $S_i^*$ in the long-range potential equation, the source term $S_i$ which produces the same solvent and solute behavior is different from $S_i^*$ since it must embed the combined effect of $\phi$ in the boundary condition on $c^{ph}$.

### 4.1 Microscopic solution

Problem (17) has been analysed in section 3.2.1 and its solution is not affected by the short-range potential hypothesis. Systems (18) and (39) are solved within the same microscopic geometry described in section 3.2.1, containing a solid circular inclusion of radius 0.35. Fields $u_i^{ph}$, $c^{ph}$ and $T$ are shown in figure 8, for $\theta = 0$, $C_0 = 1$ in equation (30). As observed in section 3.2.1. The distribution of $u_i^{ph}$ (figure 8a,b) exhibit maxima at the solid inclusion surface $\partial \mathbb{M}$ since $u_i^{ph}$ represents the velocity contribution due to the diffusio-phoretic mechanisms in the vicinity of the membrane walls (figure 8c). The chemical slip represents the only inhomogeneous source term in the problem for $u_i^{ph}$. Since the gradient of $c^{ph}$ is symmetric with respect to the vertical axis $x_n$, $u_n^{ph}$ and $u_t^{ph}$ are symmetric to the same axis. The solute-related quantity $c^{ph}$ depends on $S_i$ and represents the solute concentration variation at the membrane walls $\partial \mathbb{M}$ imposed via $S_i$. A change in $S_i$ can modify the symmetry axis of $c^{ph}$ (figure 8e) implying a rotation of the maxima in $u_i^{ph}$ and a consequent deviation of the velocity principal direction with respect to the axis $x_n$. These variations cannot be noticed in $T$, which does not depend on $S_i$ and exhibits a behavior analogous to the long-range case.

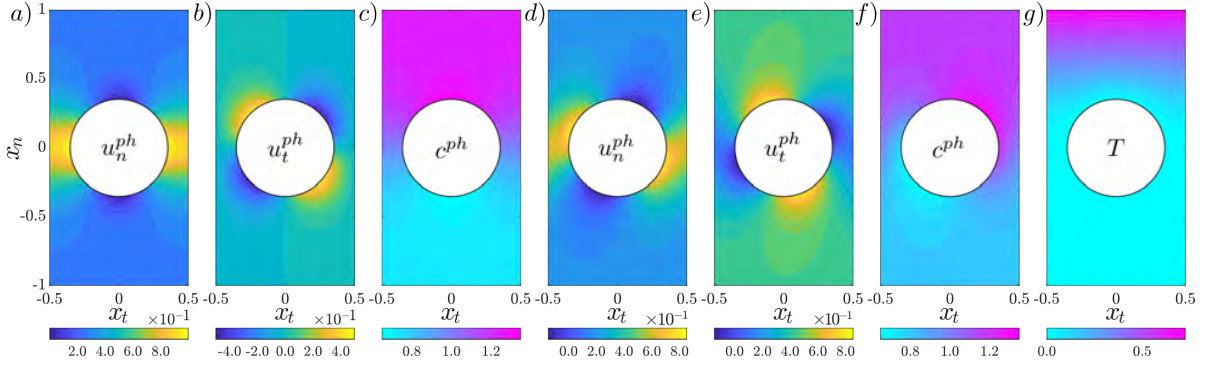

Figure 8: Contours of the microscopic fluid-flow related fields $u_n^{ph}, u_t^{ph}$ and concentration-related fields $c^{ph}, T$ in the case of short-range potential around a circular, solid inclusion for $\theta = 0$ ($a - c$) and $\theta = \pi/4$ ($d - f$) and $C_0 = 1$ in equation (30).

### 4.2 Macroscopic solution and comparisons with fully-solved simulations

As previously done for the long-range potential model, in the present section we compare the short-range potential macroscopic model to full-scale simulations of equations (31) for the same flow configuration of §3 (cf. figure 9a,b). We consider a membrane whose physio-chemical properties are described by $S_i$ with $\theta = \pi/4$. As evinced by figure 9a,b, the fluid flows from right to left and exhibits a non-negligible transversal velocity component at the membrane, induced by a concentration gradient rotated of $\theta$ with respect to the membrane centerline. While the velocity field is continuous across the membrane (9c,d), we observe a sharp concentration jump (9e). In figure 9c-e, the comparison between the average full-scale velocity components, the solute concentration (black symbols) and the macroscopic solution (colored lines) shows a good agreement.

Therefore, our homogeneous model faithfully reproduces phoretic mechanisms within the membrane in a large range of conditions, from the presence of a chemical interaction potential that varies at the scale of the elementary cell to the case in which these interactions are confined in the very vicinity of the solid walls. We now conclude our analysis by showing a simple application in which the homogenized model.

### 4.3 Towards a rational design of minimal phoretic pumps within microchannels

The generation of a net flow rate and recirculation regions in the inertialess regime without moving parts in microfluidics is a problem of large interest, see e.g. Amselem et al. [2023]. As a matter of fact, phoretic mechanisms can be utilized to generate a net flow rate in the absence of valves and moving parts Tan et al. [2019]. Here, we present a simple design of a minimal phoretic pump to guide an additional passive, inert, solute concentration field $c^*$, similar to the experimental setup of Lee et al. [2014] where an inert solute was employed to visualize the developing flow field. In our case, we consider a simple geometry that consists of a circular membrane immersed in a rectangular two-dimensional channel, which serves also as a further validation test for the model. We choose an $\epsilon$ equal to 0.025. When $\psi = 0$, on the $m$-th solid inclusion, we have

$$S_i n_i = \frac{1}{\epsilon}\left[cos(\frac{m\pi}{20})\left(x + \frac{cos(\frac{n\pi}{20})}{\pi}\right) + sin(\frac{m\pi}{20})\left(y + \frac{sin(\frac{m\pi}{20})}{\pi}\right)\right] + 1 \tag{40}$$

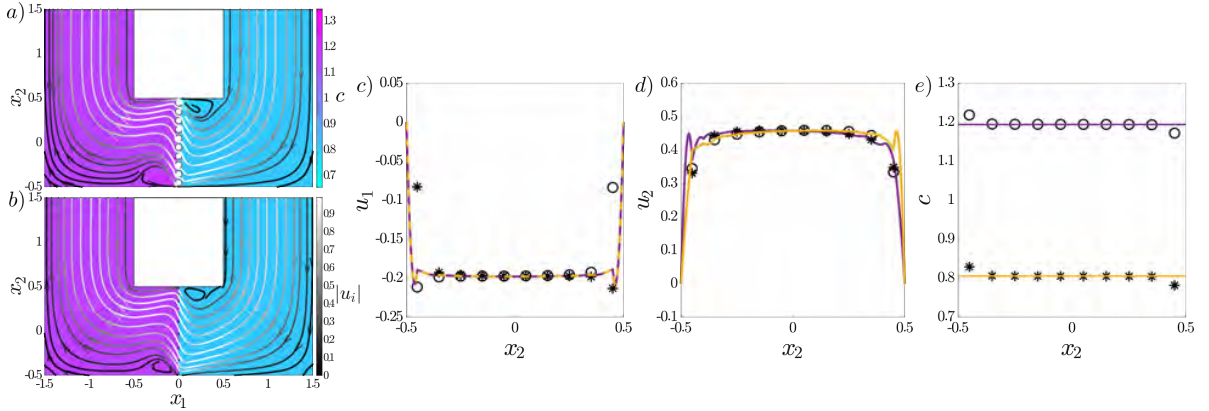

Figure 9: $\theta = \pi/4$ and $\epsilon = 0.1$. (a,b) Full-scale (panel $a$) and macroscopic ($b$) solutions in the U-shaped channel in the case of short-range potential. Contours show the solute concentration values, while streamline colours represent the local velocity magnitude. (c,d,e) Comparison between the full-scale averaged fields (black asterisks and circles for the left and right-hand sides of the membrane, respectively) and macroscopic (yellow and purple for the left- and right-hand sides of the membrane, respectively) velocity components and solute concentration on the membrane.

if $m \in [1, 6]$, $[16, 26]$ or $[36, 40]$ while, for the other values of $m$, $S_i n_i = 0$. These intervals characterize two regions (leftmost and rightmost portions of the circle) where the source is active and the other two (uppermost and lowermost regions of the circle), where the source term is zero. We note that $m = 1$ corresponds to the leftmost inclusion in figure 10a and $m$ increases counterclockwise. At the center of the circular membrane, a point-wise inlet releases continuously in time a constant concentration of $c^* = 1$, which is then advected by the steady $(u_1, u_2)$ fluid flow field. We assume that the Péclet number of the inert species is $Pe_L^* = \frac{U^{\circ}L}{D^*} = 1000$ and use the central approximation provided in Zampogna et al. [2022] to compute the equivalent diffusivity $\bar{T}^* = 0.795$ on the homogeneous interface. By rotating the membrane of an angle $\psi \in [0, \pi/2]$ we can generate a net fluid flow or recirculating patterns along the channel, see figure 10. Recirculating patterns are of utmost interest within the low Reynolds number regime since they can enable controlled microfluidic mixing and chemical reactions [Ault et al., 2018, Teng et al., 2023]. In this case, recirculations are generated without the need for pumps, valves and flexible walls or elements. The homogenized model is suitable for a fine-tuning of the properties of the membrane and a subsequent inverse design approach to retrieve these values, see e.g. Ledda et al. [2021] for the hydrodynamic case.

In addition to these considerations, we include in the analysis a tank with a small connection on the side of the channel, as shown in figure $11a - c$. We simulate the filling of the tank with a required concentration of inert solute $c^*$ by monitoring its evolution in time $\tau = t/T^{\circ} \in [0, 100]$, where $T^{\circ}$ is the macroscopic time scale defined above as $T^{\circ} = L/U^{\circ}$ for a range of $\theta$. The flow rate and concentration at the tank corner marked in red are shown in figure 11e-f. By changing the direction of the source term on the membrane we are not only creating a phoretic pump, but also we can selectively guide the inert concentration to achieve a desired local concentration at some point in the domain. The full-scale and macroscopic simulations agree well on these trends. This final application demonstrates how the homogenization technique proposed in this paper can be used even in the presence of macroscopic unsteadiness and a third, inert, species. The model allows for a cost-effective computation of phoretic flows across thin, porous membranes, both in the case of long- and short-range potentials. An indirect computational advantage of this technique resides in the absence of stiff regions in the macroscopic simulation: indeed in the full-scale simulation the local grid refinement on each solid inclusion must account for the steep descent of the potential, which may generate grids with high expansion ratios. Moreover, the method allows for the exploration of various geometry and source/potential configurations by only modifying terms in the microscopic problems, offering interesting perspectives for membrane design applications.

## 5 Conclusion and outlooks

We developed a homogeneous model to describe the diffusio-phoretic transport across micro-structured surfaces, fully permeable to the solute. The model is fully closed, i.e., it does not rely on any fitting parameter but the shape of the potential $\phi$ once chemical and hydrodynamic properties are known, and accounts for the presence of the micro-structure via a macroscopic condition on the variables imposed on a smooth single-scale interface without thickness between two fluid regions. As a matter of fact, quantities across the membrane are discontinuous due to chemical interactions at the pore scale.

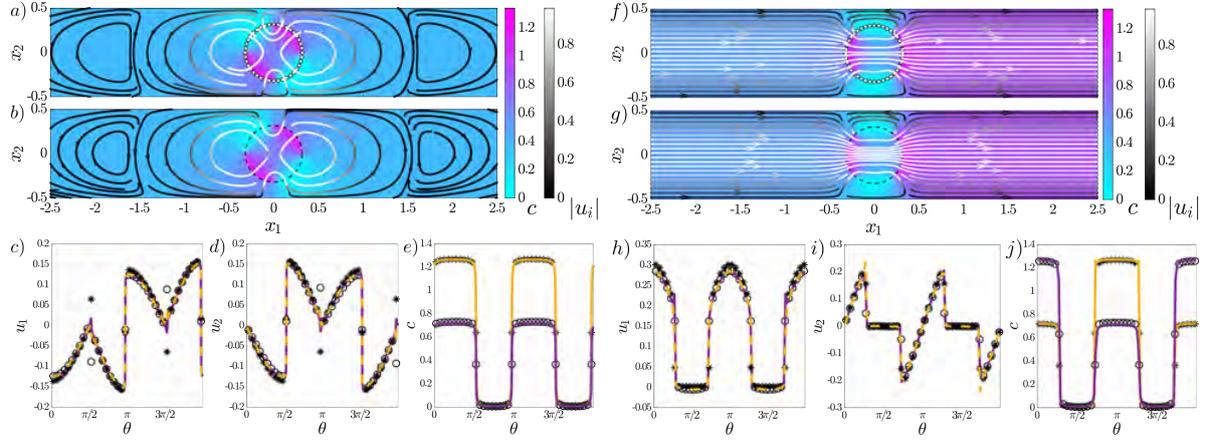

Figure 10: $(a-e)$ $\theta = 0$ and $(f-j)$ $\theta = \pi/2$. Panels $a, b$ and $f, g$: full-scale and macroscopic concentration $c$ contours and streamlines coloured by velocity magnitude, respectively. Panels $c-e$ and $h-j$: full-scale averaged (black) velocity components and concentration on the inner (asterisks) and outer (circles) and corresponding macroscopic quantities on the inner (yellow) and outer (purple) sides of the membrane.

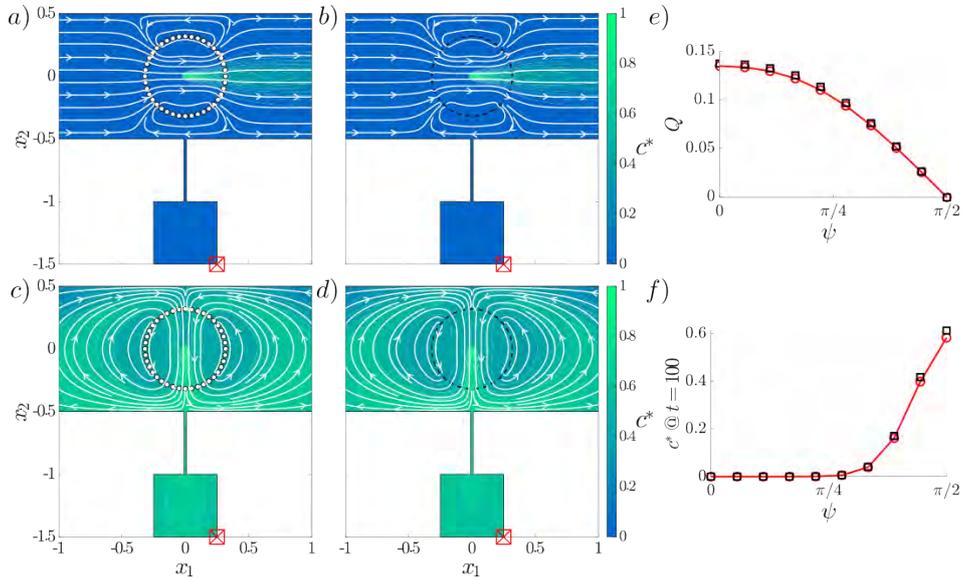

Figure 11: Channel with tank connected in the $\theta = 0$ (panels $a, b$) and $\theta = \pi/2$ (panels $c, d$) cases. Panels $a - d$ show iso-levels of inert concentration $c^*$ in the full scale $(a, c)$ and macroscopic $(b, d)$ solution at $\tau = 100$. Velocity streamlines in white. The red cross indicates the location of the concentration probe used in panel $f$. Panels $e$ and $f$ show the flow rate $Q = \int_{-0.5}^{0.5} u_i n_i dx_2$ and the inert concentration at the probe for $\tau = 100$, respectively.

We considered and validated against full-scale simulations two different levels of approximation at the microscopic scale, where the chemical potential variations (*i*) propagate into the elementary cell of the membrane and (*ii*) are felt only in the vicinity of the solid pores, allowing for a boundary condition-based description. These characterizations of physically-enabled flows in the vicinity of solid interacting walls have already been employed in the literature, e.g. by Marbach et al. [2017] and Michelin and Lauga [2014], respectively. The short-scale approximation appears as a viable method to consider the leading order chemical interactions in the vicinity of the solid walls.

From a macroscopic point of view, the velocity and the concentration are described by a linear combination of the upward and downward stress tensors and solute fluxes, and some source terms which represent the upscaled phoretic contribution to the flow, deriving from the solute-solvent-membrane interactions. Quantities present in the macroscopic condition are the averaged entries of tensors retrieved by the solution of Stokes-Smoluchowski problems with a potential source term, within a periodic microscopic domain. The phoretic contributions to the flow macroscopically result in a net flow and concentration source at the membrane, which depends on the chemical interaction on the solid boundaries of the microscopic inclusions composing the membrane.

The model shows a high degree of generality as concerns (*i*) the microscopic topology of the membrane (only periodicity along the tangential to the surface direction is imposed) and (*ii*) the macroscopic shape of the membrane (whose unique constraint is the separation of scales $l/L = \epsilon \ll 1$). Furthermore, the condition applies to generic flow configurations involving chemical and electrical interactions embedded within a specified interaction potential as well as to different chemical interactions between solid boundaries and fluid flow, thus suggesting promising applications in the modelling of phoretic mechanisms such as pumping and recirculating effects for microfluidics systems, as discussed in a dedicated example.

It is worth mentioning some limits of the model here. First, the model applies when the pore size is such that continuum Stokes-Smoluchowski equations are valid, (cf., for instance, filtration processes such as industrial wastewater treatment, and water reuse involving membranes whose pores diameter is larger than a few nanometers [Kirby, 2010]). Important biological processes happen at smaller scales, across sub-nanometric pores (cf., for instance, Verkman [2012]). Indeed, membranes are typically *geometrically*-selective, i.e. they do not allow solute particles and ions beyond a certain size to pass through the membrane. Our model does not contain such geometrical ingredient, resulting in a partial membrane permeability to the solute. For this reason, the accumulation of chemical species on the membrane (i.e. fouling and concentration polarization) is not currently modelled in the proposed framework. This ingredient involves the introduction of more refined microscopic descriptions to account for confinement effects [Gravelle et al., 2013]. These microscale considerations can be thus averaged within a homogenized description to obtain a macroscopic model coupled with the employed macroscopic continuum equations

Second, the model here presented is developed only for one transported species. However, in the limit of non-interacting diluted solutes of different species, the model can be straightforwardly generalized to several species thanks to the linear superposition principle. This task has partially been fulfilled in §4.3, when an inert[1] species was advected by the diffusio-phoretic flow. In the long-range case, we would have

$$u_i^\pm = \epsilon \overline{M}_{ij}^{\mathbb{C}^\pm} \Sigma_{jk}^\pm n_k + \epsilon \overline{N}_{ij}^{\mathbb{C}^\pm} \Sigma_{jk}^\mp n_k +$$
$$+ \epsilon^2 \sum_{n=1}^{n_{sp}} \left( \overline{A}_i^{(n),\mathbb{C}^\pm} F_k^{(n),\pm} n_k + \epsilon \overline{B}_i^{(n),\mathbb{C}^\pm} F_k^{(n),\mp} n_k + \epsilon \overline{\alpha}_i^{(n),\mathbb{C}^\pm} \right), \quad (41)$$

$$c^{(n),\pm} = \epsilon \left( \overline{T}^{(n),\mathbb{C}^\pm} F_k^{(n),\mp} n_k + \overline{Y}^{(n),\mathbb{C}^\pm} F_k^{(n),\pm} n_k + \overline{\gamma}^{(n),\mathbb{C}^\pm} \right), \quad (42)$$

and in the short-range case:

$$u_i^\pm = \epsilon \overline{M}_{ij}^{\mathbb{C}^\pm} \Sigma_{jk}^\pm n_k + \epsilon \overline{N}_{ij}^{\mathbb{C}^\pm} \Sigma_{jk}^\mp n_k + \epsilon^2 \sum_{n=1}^{n_{sp}} \left( \epsilon \overline{u_i^{ph}}^{(n),\mathbb{C}^\pm} \right), \quad (43)$$

$$c^{(n),\pm} = \epsilon \left( \overline{T}^{(n),\mathbb{C}^\pm} F_k^{(n),\mp} n_k + \overline{Y}^{(n),\mathbb{C}^\pm} F_k^{(n),\pm} n_k + \overline{c^{ph}}^{(n),\mathbb{C}^\pm} \right), \quad (44)$$

where each term $\cdot^{(n)}$ stems from the previously derived Smoluchowski-like problems for each chemical potential and boundary conditions associated with each solute.

---

[1] i.e. non interacting neither with other solutes nor with the membrane

A further extension, to close the gap toward modelling of actual complex biological systems such as the flow through the retinal epithelium [Dvoriashyna et al., 2020], would be to include the chemical reactions between different solutes within a similar general, rigorous and closed model. For example, Boso and Battiato [2013] performed homogenization of multi-component reactive species in bulk porous media.

Third, the model assumes that convective terms within the governing equations can be neglected, i.e. $Re_\ell, Pe_\ell \ll 1$. Further developments could relax the hypothesis by including an Oseen-like approximation [Zampogna et al., 2016, Wittkowski et al., 2024] to extend the range of validity of the current model toward larger scales, or different solute-solvent interactions regimes.

In summary, we have proposed a rigorous and closed model to describe transport across phoretic membranes. In addition to the good predictive power combined with significantly reduced computational times, the model is potentially suitable for optimization and inverse design procedures, see Ledda et al. [2021]. We envisage this model to be employed as a building block toward an improved physical understanding of interfacial transport across biological and engineered membranes for innovative applications, as well as for the actual design of phoretic systems for net transport and mixing of solutes within microfluidic devices and beyond.

## Declaration of interests

The authors report no conflict of interest.

## Funding

This work was supported by the Swiss National Science Foundation (grant no. PZ00P2_193180) and by the Italian Ministry of the University and Research via the "Rita Levi Montalcini" grant to GAZ.

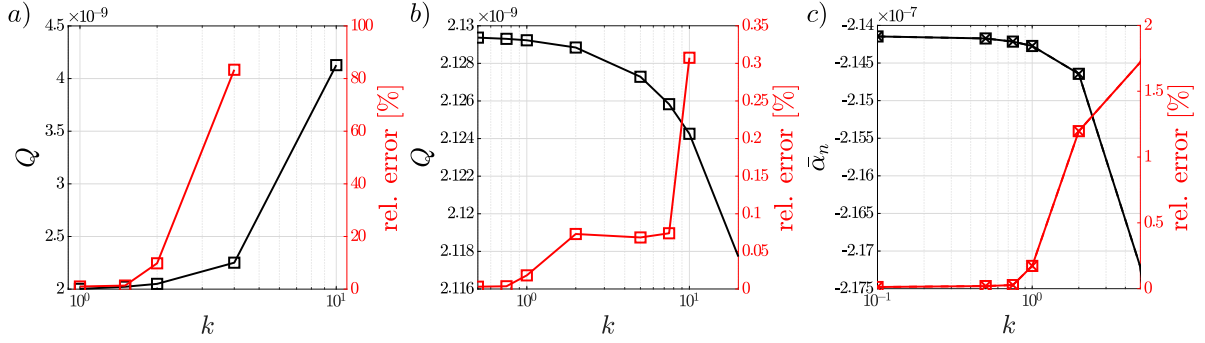

Figure 12: Flow rate $Q$ (black lines) evolving in the U-shaped pipe in the full-scale simulation (panel $a$) and macroscopic simulation ($b$) as a function of the mesh factor $k$. The corresponding relative error is shown on the left y-axis (red). Average values of $\alpha_n$ (black) sampled on the top (full line) and bottom (dashes) lines in the microscopic simulation as a function of the mesh factor $k$. Corresponding relative errors are shown on the left y-axis (red).

## A  Mesh independence study

In this Section, we assess the robustness of the results with respect to the mesh size. As a test case, we consider the long-range case. As a matter of fact, the presence of a boundary layer in the vicinity of the solid walls due to the functional form of potential $\phi$ is computationally more challenging than the short-range case, where the interaction appears as a boundary condition. The parameters of the initial grid are

- for the full-scale computation (the reference setup is shown in figure 7a): overall maximum mesh size of $0.05L$, maximum element size near the solid inclusions of $0.00038L$ with first cell height of the prismatic layers of $6.25 \times 10^{-6} L$;
- for the macroscopic computation (the reference setup is shown in figure 7b): overall maximum mesh size of $0.1L$, maximum element size near the solid inclusions of $0.001L$;
- for the microscopic computation (corresponding to figure 3): overall maximum mesh size of $0.02\ell$, maximum element size near the solid inclusions of $0.001\ell$.

All these mesh parameters but the overall sizing were perturbed by multiplying them for a factor $k$ and computing a solution for each $k$. Figure 12 shows that for $k = 1$ in all cases we can expect an error of order

- 1.0% on the flow rate $Q$ evolving in the system for the full-scale simulation;
- 0.018% on the flow rate $Q$ evolving in the system for the macroscopic simulation;
- 0.17% on the values of $\bar{\alpha}_n$ for the microscopic simulation.

Meshes with $k = 1$ have been employed for all calculations performed in this work.